\documentclass[12pt]{article}
\usepackage[dvips]{graphicx}
\usepackage{pdproc}

  \makeatletter 
  \def\@cite#1{[#1]} 
  \makeatother    
\begin{document}

\renewcommand{\thefootnote}{\alph{footnote}}

\title{
SuperWIMP Dark Matter in Supergravity with a Gravitino 
LSP \footnote{Presented by Shufang Su at 
SUSY2004, June 17-23, 2004, Tsukuba, Japan.}}

\author{Jonathan L.~Feng$^1$, Shufang Su$^2$, Fumihiro Takayama$^1$}
\address{$^1$Department of Physics and Astronomy, University of
California, Irvine, CA 92697, USA\\
$^2$Department of Physics, University of Arizona,
Tucson, AZ 85721, USA}



\vspace*{- 0.25 in}
\abstract{
We investigate the superWIMP scenario in the framework of supersymmetry,
in which the lightest supersymmetric particle (LSP) is a stable gravitino.
We consider slepton, sneutrino or neutralino being the next-lightest 
supersymmetric particle (NLSP), and determine 
what superpartner masses are viable,
applying cosmic-microwave background (CMB) 
and electromagnetic and hadronic Big-Bang Nucleosythesis (BBN) constraints.
}

\normalsize\baselineskip=15pt

\section{Introduction}

SuperWIMPs, superweakly interacting massive particles produced in the
late decays of weakly-interacting massive particles (WIMPs):
${\rm WIMP} \rightarrow {\rm superWIMP} + {\rm SM\ particle}$, 
are promising non-baryonic dark matter
candidates~\cite{Feng:2003xh}. 
The relic density of superWIMP is related to the relic 
density of WIMP via 
\begin{equation}
\Omega_{\rm SWIMP}= \frac{m_{\rm SWIMP}}{m_{\rm WIMP}} 
\Omega_{\rm WIMP}.
\end{equation}
For $m_{\rm SWIMP}\sim m_{\rm WIMP} \sim$ Electroweak scale,
superWIMP inherits the relic density of WIMP, which is naturally of
the right magnitude near the observed value. 
Well-motivated superWIMP
candidates are the gravitinos in supersymmetric
models
~\cite{Feng:2003xh,threebody1,Ellis:2003dn,Wang:2004ib} 
and the first excited gravitons in universal
extra dimension models~\cite{Feng:2003xh,Feng:2003nr}.
In this talk, I will focus on the gravitino LSP in the 
supersymmetric models as a concrete example.

In the supersymmetric framework with the  gravitino LSP being the superWIMP, 
the NLSP, which is the WIMP, could be charged slepton $\tilde{l}$, 
sneutrino $\tilde{\nu}$, or neutralino/chargino $\chi$.  
The dominate decay modes for the NLSPs are
\begin{equation}
\tilde{l} \rightarrow \tilde{G} + {l};\ \ \
\tilde{\nu} \rightarrow \tilde{G} + \nu;  \ \ \ 
\chi \rightarrow \tilde{G} + \gamma/ Z/ W/ h, 
\label{eq:decay}
\end{equation}
which occur at $t \sim 10^4 - 10^8$ sec, well after BBN.  
An immediate concern, therefore, is that they might destroy
the successful light element abundance predictions of BBN.  
The  constraints on
the electromagnetic (EM) energy released in the decays of
Eq.~(\ref{eq:decay}) exclude some of the weak scale parameter space, but
leave much of it intact \cite{Feng:2003xh}.  
The BBN constraints on the hadronic energy 
release, however, is much stronger.  
For slepton and sneutrino, the hadronic decay could only occur via 
subdominant three or four body decay:
\begin{equation}
\tilde{l} \rightarrow \tilde{G} + Z/W + l^{\prime} \rightarrow 
\tilde{G}+q \bar{q}^{\prime}+l^{\prime}; \ \ \ 
\tilde{l} \rightarrow \tilde{G} + \gamma^* + l 
\rightarrow \tilde{G}+q \bar{q}+l.
\end{equation}
The branching ratio is usually smaller than ${\cal{O}}(10^{-3})$
in most of the parameter space.  
For neutralino/chargino, however, the hadronic decay branching ratio is 
unsuppressed since $Z/W/h$ in Eq.~(\ref{eq:decay}) could decay hadronically.
Imposing the hadronic BBN constraints would disfavor 
much of the parameter spaces \cite{threebody1}.

\section{Late Time Decay and BBN Constraints}
The overall success of standard BBN places severe constraints on
energy produced by particles decaying after BBN.  
In our analysis, we use the BBN constraints on EM energy release in 
Ref.~\cite{Cyburt:2002uv} and the latest analysis on 
both EM and hadronic energy releases 
in Ref.~\cite{Kawasaki:2004yh}.
For time $t>10^4$ sec, the BBN constraints are, to a good
approximation, constraints on energy injection $\xi_{\rm i}$: 
$\xi_{\rm {i}} \equiv 
\epsilon_{\rm{i}} B_{\rm{i}} Y_{\rm{NLSP}}$, 
where $\rm{i} = \rm{EM}, \rm{had}$.  
Here $B_{\rm{i}}$ is the branching fraction
into EM/hadronic components, and $\epsilon_{\rm{i}}$ is the
EM/hadronic energy released in each NLSP decay.  
$Y_{\rm{NLSP}} \equiv
n_{\rm{NLSP}}/n_{\gamma}^{\rm{BG}}$ is the NLSP number density
just before NLSP decay, normalized to the background photon number
density $n_{\gamma}^{\rm{BG}} = 2 \zeta(3) T^3 /
\pi^2$.


In evaluating the constraints, we use two different  approaches for 
$\Omega_{\rm SWIMP}$:
\begin{itemize}
\item{\bf Approach I:}  SuperWIMP gravitinos make up all of the
non-baryonic dark matter, with
$\Omega_{\tilde{G}} \simeq 0.23$.
\item{\bf Approach II:} The NLSP
freezes out with its thermal relic density
$\Omega_{{\rm NLSP}}^{{\rm th}}$.  The superWIMP gravitino density
is then $\Omega_{\tilde{G}} = (m_{\tilde{G}}/m_{{\rm NLSP}})
\Omega_{{\rm NLSP}}^{{\rm th}}$.  In this approach, the superWIMP  
gravitino density may be low and even insignificant cosmologically. 
\end{itemize}
These approaches differ significantly.
Low masses
are excluded in the former case, while high masses are disfavored in
the latter. 
This difference has obviously
important implications for collider searches for new physics.

\section{Results}
{\bf Approach I: Fix $\Omega_{\tilde{G}}=0.23$.}

\begin{figure}
\begin{center}
\includegraphics*[width=2.25 in]{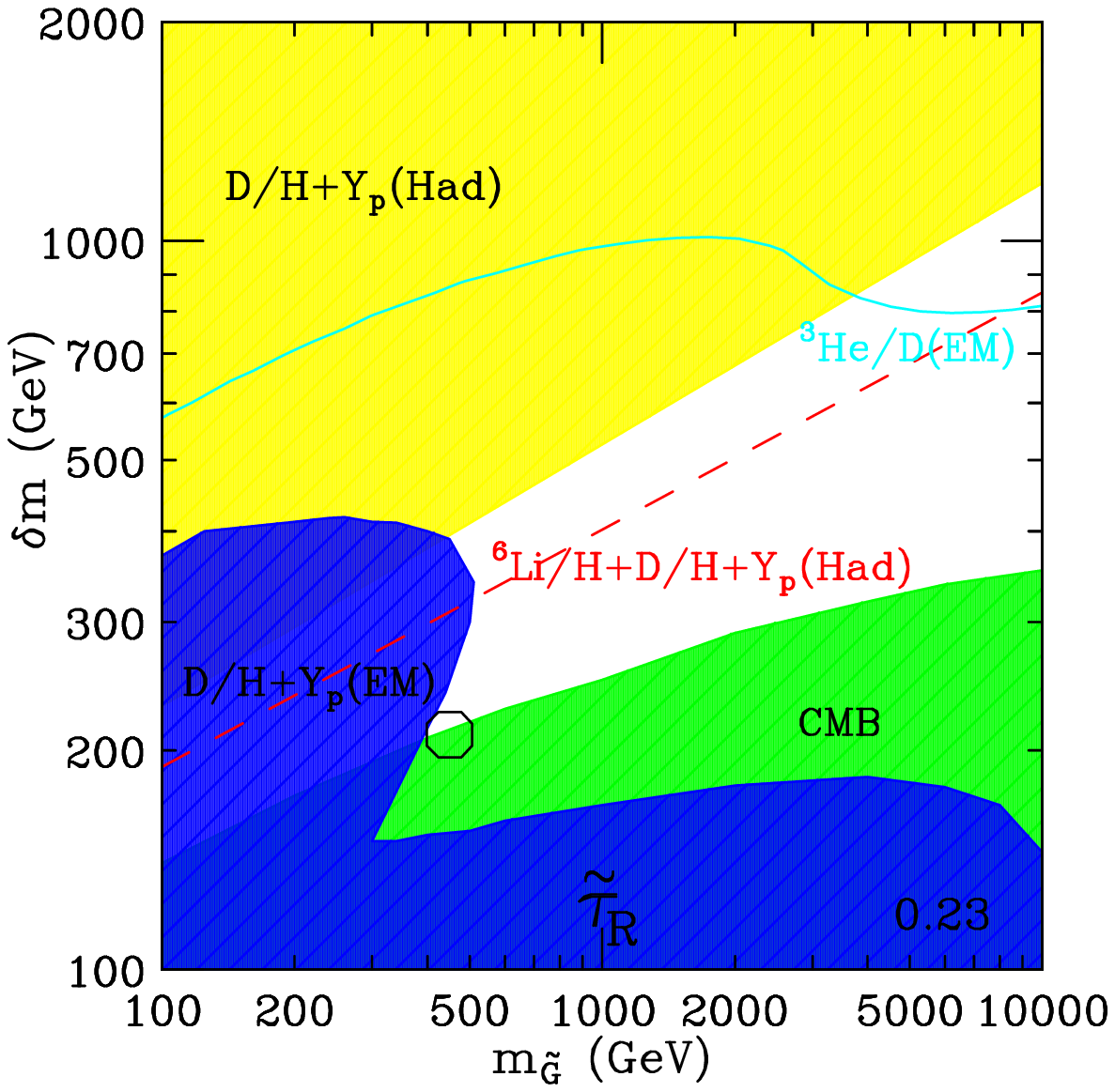}
\includegraphics*[width=2.25 in]{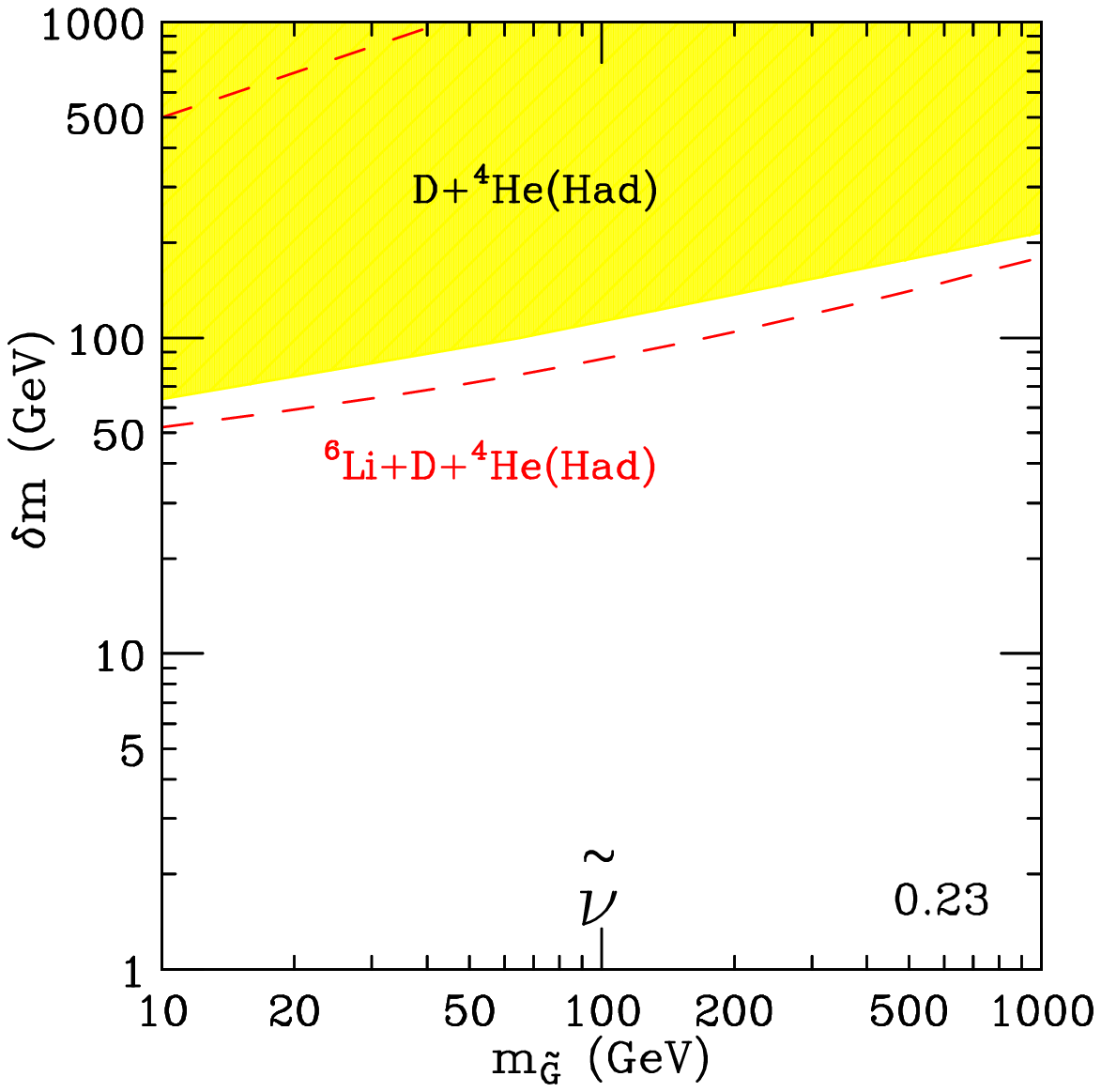}
\end{center}
\caption{Allowed and disfavored regions of the $(m_{\tilde{G}}, \delta
m)$ plane, where $\delta m \equiv m_{{\rm NLSP}} - m_{\tilde{G}} -
m_Z$, for (left) $\tilde{\tau}_R$ and (right) $\tilde{\nu}$ NLSPs.
The shaded regions are disfavored by CMB, EM BBN, and hadronic BBN
constraints, as indicated.  The lines correspond to EM constraints
from ${}^3$He/D (solid) and hadronic constraints from ${}^6$Li/H
(dashed).  
The circle indicates the best fit region where the $^7$Li
discrepancy is resolved.  We have assumed
$\Omega_{\rm SWIMP} = 0.23$ and $\epsilon_{{\rm had}}=\frac{1}{3}
(m_{{\rm NLSP}}-m_{\tilde{G}})$.  
}
\label{fig:summary} 
\end{figure}

Fig.~\ref{fig:summary} shows the allowed and disfavored regions of the
$(m_{\tilde{G}}, \delta m)$ plane for the two NLSP cases.  
The neutralino/chargino
parameters are chosen to be $M_1=2 m_{{\rm NLSP}}$, $M_2=\mu=4
m_{{\rm NLSP}}$, and $\tan\beta=10$.
The CMB
$\mu$ distortion constraint, the BBN constraint of
Ref.~\cite{Cyburt:2002uv} and 
Ref.~\cite{Kawasaki:2004yh} are included.  
For the $\tilde{\tau}_R$
case,  the hadronic constraint is extremely important --- it disfavors a
large and natural part of parameter space that would otherwise be
allowed.
Even after all of these constraints, however, the unshaded area in the
region $m_{\tilde{G}} \geq 200~{\rm GeV}$ and $200~{\rm GeV} \leq 
\delta m \leq
1500~{\rm GeV}$ still remains viable.  The $\tilde{\tau}_R$ mass must be
above 500 GeV, which  is within reach of the LHC.  

The sneutrino NLSP case is also shown in Fig.~\ref{fig:summary}.
Hadronic BBN constraints from the D and $^4$He abundances only
disfavor $\delta m \geq$ 100 GeV. Even including the stronger but more
speculative ${}^6$Li/H constraint, there is still a large region of
$(m_{\tilde{G}}, \delta m)$ that is allowed.  

The neutralino/chargino NLSP can not be realized in this approach, due to the 
unsuppressed hadronic decay branching ratio.
\\

{\bf Approach II: 
$\Omega_{\tilde{G}}=(m_{\tilde{G}}/m_{\rm NLSP})\Omega_{\rm NLSP}$.}

\begin{figure}
\begin{center}
\includegraphics*[width=2.25 in]{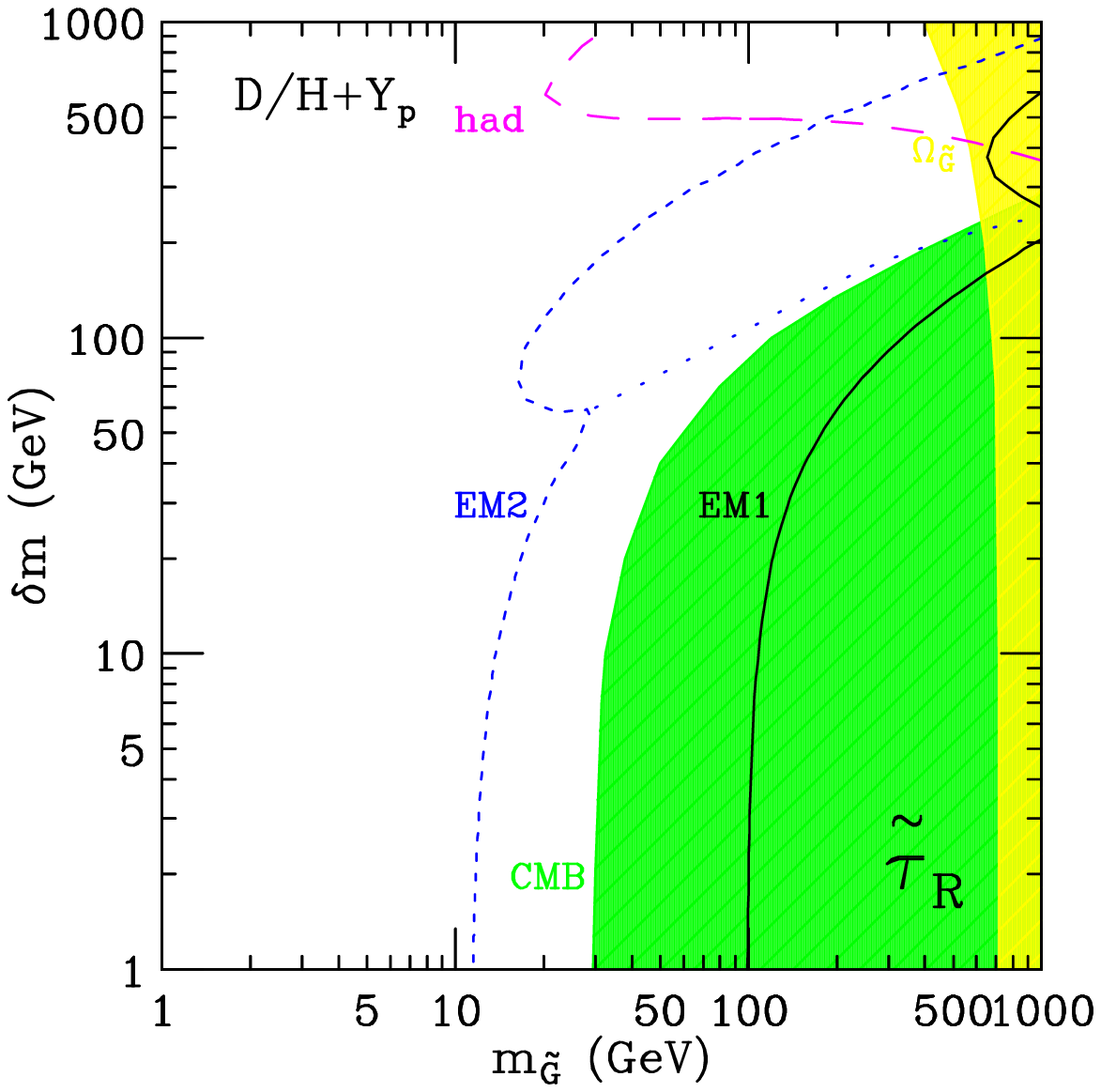}
\includegraphics*[width=2.25 in]{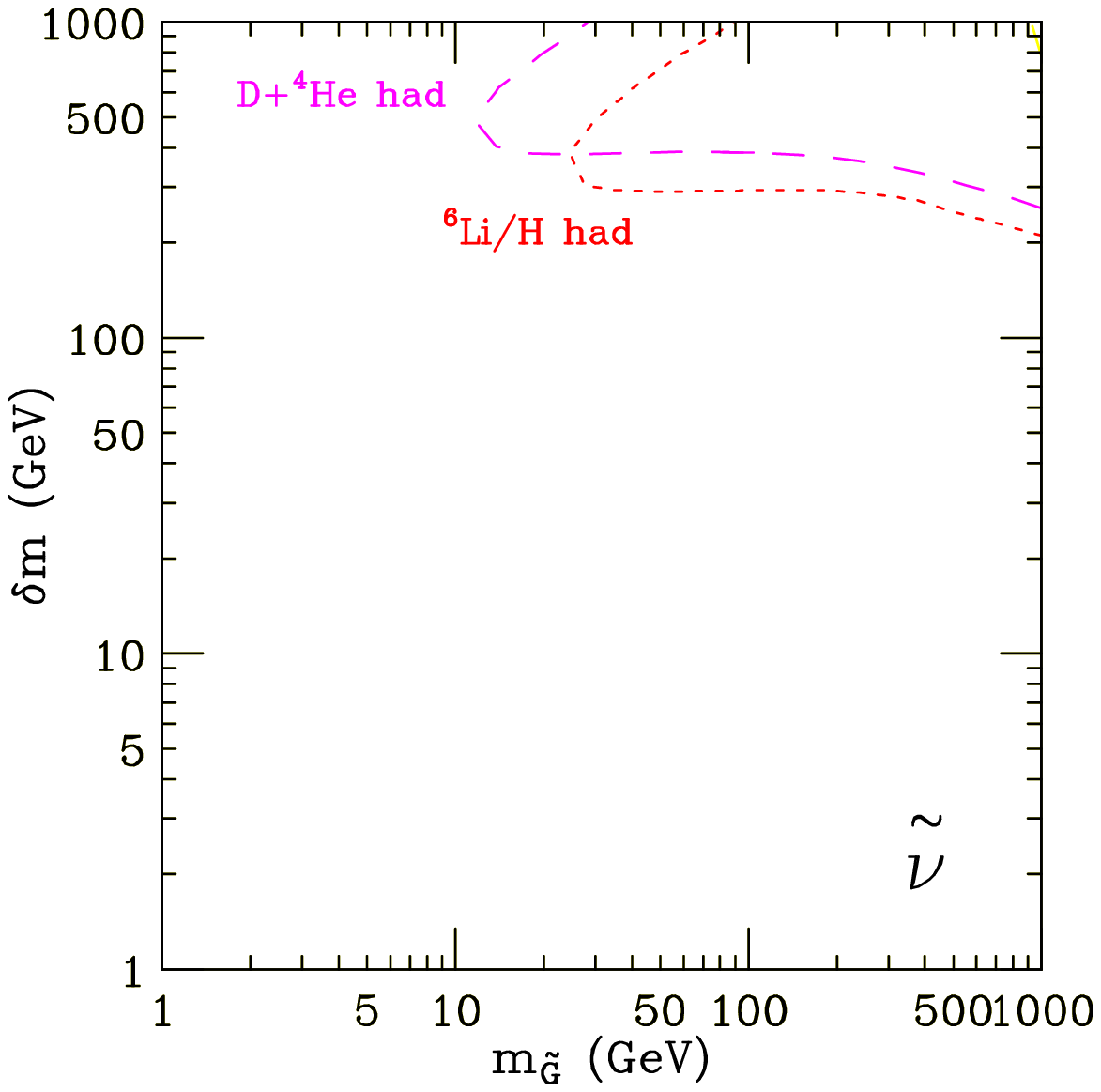}
\end{center}
\caption{
Excluded and allowed regions of the $(m_{\tilde{G}}, \delta m )$ 
parameter space in
the gravitino LSP scenario for (left) $\tilde{\tau}_R$ 
and (right) $\tilde{\nu}$ NLSPs, assuming a 
$\tilde{\tau}_R$ or $\tilde{\nu}$ NLSP that
freezes out with thermal relic density given by Eq.~(\ref{omega}). The
light (yellow) shaded region is excluded by the overclosure constraint
$\Omega_{\tilde{G}} < 0.23$, and the medium (green) shaded region
is excluded by the absence of CMB $\mu$ distortions.  BBN is sensitive
to the regions to the right of the labeled contours.  ``EM1'' contour
indicate the regions
probed by D and $^4$He in Ref.~\protect\cite{Cyburt:2002uv};
``EM2'' and ``had'' contour indicate the 
more stringent constraints of Ref.~\protect\cite{Kawasaki:2004yh}.
The dotted line denotes the region
where cancellation between D destruction and creation via late time EM
injection is possible, while ${}^7$Li is
reduced to the observed value by the late NLSP
decays~\protect\cite{Cyburt:2002uv}.  
}
\label{fig:stausneu}  
\end{figure}

The results are presented in Fig.~\ref{fig:stausneu} 
for right-handed stau and sneutrino, where we have 
adopted a simple scaling behavior for the thermal relic density:
\begin{equation}
\Omega_{\tilde{l}_R}^{{\rm th}}\approx 0.4 \,
\left[\frac{m_{\tilde{l}_R}}{\rm TeV}\right]^2; \ \ \ 
\Omega_{\tilde{\nu}}^{{\rm th}} \approx 0.12 \,
\left[\frac{m_{\tilde{\nu}}}{\rm TeV}\right]^2 \ . 
\label{omega}
\end{equation} 

The shaded regions are excluded by the overclosure constraints and
the absence of CMB $\mu$ distortions.
The BBN sensitivity contours divided into those 
from Ref.~\cite{Cyburt:2002uv} (EM1) and from Ref.~\cite{Kawasaki:2004yh}
(EM2, had).
The strength
of the EM constraints in constraining gravitino LSP parameter space
depends sensitively on how one interprets the BBN data.  Adopting the
more stringent EM2 contour, we find that bounds on $m_{\tilde{G}}$ are
improved by about an order of magnitude.
For relatively early decays, the hadronic
constraint is the leading constraint, 
which covers  the most natural parameter region for weak-scale
supergravity.  

Sneutrinos annihilate
through $S$-wave processes even more efficiently than sleptons.
The dark matter
density bound is therefore weaker.
The remaining constraints are therefore only the hadronic BBN
constraints.  These are stringent for early decays, that is, large
$\delta m$.  The more reliable D and $^4$He constraints disfavor
$\delta m \geq 300~{\rm GeV}$, while $^6$Li (had) is sensitive to $\delta m
\geq 200~{\rm GeV}$.  

For neutralino NLSP, the strong BBN hadronic constrains exclude most of the 
parameter space except when  the relic density of the gravitinos 
is insignificant.  For a pure Bino,
the viable parameter space is very limited, mostly for $m_{\tilde{G}}$ below 
about 100 MeV. 




\section{Summary and Outlook}
In this work, we have determined the viability of superWIMP dark matter 
in the 
framework of supersymmetric  scenarios in which the gravitino is the LSP.  
We found that the hadronic BBN constraints, previously neglected, are
extremely important, providing the most stringent limits in natural
regions of parameter space \cite{threebody1}.

The gravitino LSP scenario opens up many connections between particle
physics and cosmology.   The reduced Planck 
mass ~\cite{Buchmuller:2004rq,Feng:2004gn} could be measured by 
trapping a charged slepton and observe its consequent decay. 
A more detailed analysis
of the trapping of sleptons at future colliders is now under
study~\cite{FMS}.

\section{Acknowledgments}

The work of JLF was supported in part by
National Science Foundation CAREER Award PHY--0239817, and in part by
the Alfred P.~Sloan Foundation.

\bibliographystyle{plain}

\end{document}